\RequirePackage{lineno}
\documentclass[aps,prl,twocolumn,showpacs,superscriptaddress]{revtex4}  
\usepackage{graphicx}
\usepackage{dcolumn}
\usepackage{bm}
\usepackage{color}
\usepackage{amssymb}
\usepackage{epstopdf}

\def\missET {{\not\!\!E_T}}
\def\missETvec {{\not\!\!\vec{E}_T}}

\newcommand{\unit}[1]{\,\mathrm{#1}}  
\begin{document}
\hspace{5.2in} \mbox{Fermilab-PUB-10-365-E}

\title{\boldmath Improved Search for a Higgs Boson Produced in Association with $Z \rightarrow \ell^{+} \ell^{-}$ in $p\bar{p}$ Collisions at $\sqrt{s} = 1.96\unit{TeV}$ \unboldmath} 

\affiliation{Institute of Physics, Academia Sinica, Taipei, Taiwan 11529, Republic of China} 
\affiliation{Argonne National Laboratory, Argonne, Illinois 60439, USA} 
\affiliation{University of Athens, 157 71 Athens, Greece} 
\affiliation{Institut de Fisica d'Altes Energies, Universitat Autonoma de Barcelona, E-08193, Bellaterra (Barcelona), Spain} 
\affiliation{Baylor University, Waco, Texas 76798, USA} 
\affiliation{Istituto Nazionale di Fisica Nucleare Bologna, $^{bb}$University of Bologna, I-40127 Bologna, Italy} 
\affiliation{Brandeis University, Waltham, Massachusetts 02254, USA} 
\affiliation{University of California, Davis, Davis, California 95616, USA} 
\affiliation{University of California, Los Angeles, Los Angeles, California 90024, USA} 
\affiliation{Instituto de Fisica de Cantabria, CSIC-University of Cantabria, 39005 Santander, Spain} 
\affiliation{Carnegie Mellon University, Pittsburgh, Pennsylvania 15213, USA} 
\affiliation{Enrico Fermi Institute, University of Chicago, Chicago, Illinois 60637, USA}
\affiliation{Comenius University, 842 48 Bratislava, Slovakia; Institute of Experimental Physics, 040 01 Kosice, Slovakia} 
\affiliation{Joint Institute for Nuclear Research, RU-141980 Dubna, Russia} 
\affiliation{Duke University, Durham, North Carolina 27708, USA} 
\affiliation{Fermi National Accelerator Laboratory, Batavia, Illinois 60510, USA} 
\affiliation{University of Florida, Gainesville, Florida 32611, USA} 
\affiliation{Laboratori Nazionali di Frascati, Istituto Nazionale di Fisica Nucleare, I-00044 Frascati, Italy} 
\affiliation{University of Geneva, CH-1211 Geneva 4, Switzerland} 
\affiliation{Glasgow University, Glasgow G12 8QQ, United Kingdom} 
\affiliation{Harvard University, Cambridge, Massachusetts 02138, USA} 
\affiliation{Division of High Energy Physics, Department of Physics, University of Helsinki and Helsinki Institute of Physics, FIN-00014, Helsinki, Finland} 
\affiliation{University of Illinois, Urbana, Illinois 61801, USA} 
\affiliation{The Johns Hopkins University, Baltimore, Maryland 21218, USA} 
\affiliation{Institut f\"{u}r Experimentelle Kernphysik, Karlsruhe Institute of Technology, D-76131 Karlsruhe, Germany} 
\affiliation{Center for High Energy Physics: Kyungpook National University, Daegu 702-701, Korea; Seoul National University, Seoul 151-742, Korea; Sungkyunkwan University, Suwon 440-746, Korea; Korea Institute of Science and Technology Information, Daejeon 305-806, Korea; Chonnam National University, Gwangju 500-757, Korea; Chonbuk National University, Jeonju 561-756, Korea} 
\affiliation{Ernest Orlando Lawrence Berkeley National Laboratory, Berkeley, California 94720, USA} 
\affiliation{University of Liverpool, Liverpool L69 7ZE, United Kingdom} 
\affiliation{University College London, London WC1E 6BT, United Kingdom} 
\affiliation{Centro de Investigaciones Energeticas Medioambientales y Tecnologicas, E-28040 Madrid, Spain} 
\affiliation{Massachusetts Institute of Technology, Cambridge, Massachusetts 02139, USA} 
\affiliation{Institute of Particle Physics: McGill University, Montr\'{e}al, Qu\'{e}bec, Canada H3A~2T8; Simon Fraser University, Burnaby, British Columbia, Canada V5A~1S6; University of Toronto, Toronto, Ontario, Canada M5S~1A7; and TRIUMF, Vancouver, British Columbia, Canada V6T~2A3} 
\affiliation{University of Michigan, Ann Arbor, Michigan 48109, USA} 
\affiliation{Michigan State University, East Lansing, Michigan 48824, USA}
\affiliation{Institution for Theoretical and Experimental Physics, ITEP, Moscow 117259, Russia}
\affiliation{University of New Mexico, Albuquerque, New Mexico 87131, USA} 
\affiliation{Northwestern University, Evanston, Illinois 60208, USA} 
\affiliation{The Ohio State University, Columbus, Ohio 43210, USA} 
\affiliation{Okayama University, Okayama 700-8530, Japan} 
\affiliation{Osaka City University, Osaka 588, Japan} 
\affiliation{University of Oxford, Oxford OX1 3RH, United Kingdom} 
\affiliation{Istituto Nazionale di Fisica Nucleare, Sezione di Padova-Trento, $^{cc}$University of Padova, I-35131 Padova, Italy} 
\affiliation{LPNHE, Universite Pierre et Marie Curie/IN2P3-CNRS, UMR7585, Paris, F-75252 France} 
\affiliation{University of Pennsylvania, Philadelphia, Pennsylvania 19104, USA}
\affiliation{Istituto Nazionale di Fisica Nucleare Pisa, $^{dd}$University of Pisa, $^{ee}$University of Siena and $^{ff}$Scuola Normale Superiore, I-56127 Pisa, Italy} 
\affiliation{University of Pittsburgh, Pittsburgh, Pennsylvania 15260, USA} 
\affiliation{Purdue University, West Lafayette, Indiana 47907, USA} 
\affiliation{University of Rochester, Rochester, New York 14627, USA} 
\affiliation{The Rockefeller University, New York, New York 10065, USA} 
\affiliation{Istituto Nazionale di Fisica Nucleare, Sezione di Roma 1, $^{gg}$Sapienza Universit\`{a} di Roma, I-00185 Roma, Italy} 

\affiliation{Rutgers University, Piscataway, New Jersey 08855, USA} 
\affiliation{Texas A\&M University, College Station, Texas 77843, USA} 
\affiliation{Istituto Nazionale di Fisica Nucleare Trieste/Udine, I-34100 Trieste, $^{hh}$University of Trieste/Udine, I-33100 Udine, Italy} 
\affiliation{University of Tsukuba, Tsukuba, Ibaraki 305, Japan} 
\affiliation{Tufts University, Medford, Massachusetts 02155, USA} 
\affiliation{Waseda University, Tokyo 169, Japan} 
\affiliation{Wayne State University, Detroit, Michigan 48201, USA} 
\affiliation{University of Wisconsin, Madison, Wisconsin 53706, USA} 
\affiliation{Yale University, New Haven, Connecticut 06520, USA} 
\author{T.~Aaltonen}
\affiliation{Division of High Energy Physics, Department of Physics, University of Helsinki and Helsinki Institute of Physics, FIN-00014, Helsinki, Finland}
\author{B.~\'{A}lvarez~Gonz\'{a}lez$^v$}
\affiliation{Instituto de Fisica de Cantabria, CSIC-University of Cantabria, 39005 Santander, Spain}
\author{S.~Amerio}
\affiliation{Istituto Nazionale di Fisica Nucleare, Sezione di Padova-Trento, $^{cc}$University of Padova, I-35131 Padova, Italy} 

\author{D.~Amidei}
\affiliation{University of Michigan, Ann Arbor, Michigan 48109, USA}
\author{A.~Anastassov}
\affiliation{Northwestern University, Evanston, Illinois 60208, USA}
\author{A.~Annovi}
\affiliation{Laboratori Nazionali di Frascati, Istituto Nazionale di Fisica Nucleare, I-00044 Frascati, Italy}
\author{J.~Antos}
\affiliation{Comenius University, 842 48 Bratislava, Slovakia; Institute of Experimental Physics, 040 01 Kosice, Slovakia}
\author{G.~Apollinari}
\affiliation{Fermi National Accelerator Laboratory, Batavia, Illinois 60510, USA}
\author{J.A.~Appel}
\affiliation{Fermi National Accelerator Laboratory, Batavia, Illinois 60510, USA}
\author{A.~Apresyan}
\affiliation{Purdue University, West Lafayette, Indiana 47907, USA}
% ...
\collaboration{CDF Collaboration\footnote{With visitors from $^a$University of Massachusetts Amherst, Amherst, Massachusetts 01003,
$^b$Istituto Nazionale di Fisica Nucleare, Sezione di Cagliari, 09042 Monserrato (Cagliari), Italy,
$^c$University of California Irvine, Irvine, CA  92697, 
$^d$University of California Santa Barbara, Santa Barbara, CA 93106
$^e$University of California Santa Cruz, Santa Cruz, CA  95064,
$^f$CERN,CH-1211 Geneva, Switzerland,
$^g$Cornell University, Ithaca, NY  14853, 
$^h$University of Cyprus, Nicosia CY-1678, Cyprus, 
$^i$University College Dublin, Dublin 4, Ireland,
$^j$University of Fukui, Fukui City, Fukui Prefecture, Japan 910-0017,
$^k$Universidad Iberoamericana, Mexico D.F., Mexico,
$^l$Iowa State University, Ames, IA  50011,
$^m$University of Iowa, Iowa City, IA  52242,
$^n$Kinki University, Higashi-Osaka City, Japan 577-8502,
$^o$Kansas State University, Manhattan, KS 66506,
$^p$University of Manchester, Manchester M13 9PL, England,
$^q$Queen Mary, University of London, London, E1 4NS, England,
$^r$Muons, Inc., Batavia, IL 60510,
$^s$Nagasaki Institute of Applied Science, Nagasaki, Japan, 
$^t$National Research Nuclear University, Moscow, Russia,
$^u$University of Notre Dame, Notre Dame, IN 46556,
$^v$Universidad de Oviedo, E-33007 Oviedo, Spain, 
$^w$Texas Tech University, Lubbock, TX  79609, 
$^x$IFIC(CSIC-Universitat de Valencia), 56071 Valencia, Spain,
$^y$Universidad Tecnica Federico Santa Maria, 110v Valparaiso, Chile,
$^z$University of Virginia, Charlottesville, VA  22906,
$^{aa}$Yarmouk University, Irbid 211-63, Jordan,
$^{ii}$On leave from J.~Stefan Institute, Ljubljana, Slovenia, 
}}
\noaffiliation

\begin{abstract} 
We present a search for the standard model Higgs boson produced with a $Z$ boson in $4.1\,\mathrm{fb^{-1}}$ of data collected with the  CDF~II  detector at the Tevatron.
In events consistent with the decay of the Higgs boson to a bottom-quark pair and the $Z$ boson to electrons or muons, we set $95\%$ credibility level upper limits on the $ZH$ production cross section times the $H\rightarrow b \bar{b}$ branching ratio. 
Improved analysis methods enhance signal sensitivity by $20\%$ relative to previous searches beyond the gain due to the larger data sample.
At a Higgs boson mass of $115\,\mathrm{GeV}/c^2$ we set a limit of $5.9$ times the standard model value.

\end{abstract}


\maketitle 

\date{\today}

Establishing the mechanism of electroweak symmetry breaking (EWSB) is one of the outstanding issues of particle physics. 
In the standard model (SM), EWSB is mediated by a Higgs field that manifests
a particle, the as-yet-unobserved Higgs boson.
A SM Higgs boson with mass ($M_H$) below $114.4\unit{GeV}/c^2$ or with $M_H$ between $162$ and $166\unit{GeV}/c^2$ has been excluded at $95\%$ confidence level in direct searches at LEP \cite{LEP_EWWG} and the Tevatron \cite{TEV_HWW2010}.
A recently reported~\cite{:2010ar} preliminary update of the Tevatron result has expanded this 95\% exclusion range to values of $M_H$ between $158$ and $175\unit{GeV}/c^2$.

At the Tevatron, production of the Higgs boson is dominated by the direct production process $gg\to H$, and for $M_H<135$~GeV$/c^2$ the Higgs boson decays primarily to a pair of $b$~quarks, $H\to b\bar{b}$~\cite{BBMAIN}. 
The process $gg\to H\to b\bar{b}$ is overwhelmed by multi-jet background.
Associated production of a Higgs boson with a leptonically decaying $W$ or $Z$ boson yields a distinct signature for efficient selection and in turn greater sensitivity despite a significantly smaller cross section than direct production~\cite{fixBrein:2003wg,fixCiccolini:2003jy}. 
This Letter presents an improved search for the SM Higgs boson produced in association with a $Z$ boson, $ZH \to \ell^+\ell^-  b\bar{b}$ 
(where $\ell$ is either an electron, $e$, or muon, $\mu$)
using $1.96\unit{TeV}$ $p\bar{p}$ collision data corresponding to $4.1\unit{fb^{-1}}$ of integrated luminosity collected with the CDF~II detector~\cite{detector}.
This search channel is one of the most sensitive to a low-mass SM Higgs boson at the Tevatron~\cite{Aaltonen:2009dh, Aaltonen:2009jg}. 

Previous CDF efforts in this mode used 
an artificial neural network classifier (NN)~\cite{ZHllbb_2008} or a likelihood based on matrix-element probabilities (MEP)~\cite{ZHLLBBME} for signal isolation.
Here we enhance these techniques with NN-based $b$~jet discrimination~\cite{KNN,singleTopPRDsubmittedKNN} and an improved multivariate jet-energy correction to improve background rejection. 
New $Z \to e^+e^-$ selections increase the acceptance of $ZH$ signal, and new combinations of $b$~jet identifiers yield better signal sensitivity, as reflected in the expected cross section limit.
These additions improve the expected sensitivity by a factor of $1.2$ over the gain expected just from additional integrated luminosity.


The selection requirements most relevant to this analysis are discussed below; a detailed presentation can be found in \cite{Shalhout}.
We select $ZH$ candidates by first identifying a sample of events containing a $Z\rightarrow \ell^+ \ell^-$ decay.
Events are selected in real time (triggered) based on the presence of high-$p_T$ electron and muon~\cite{coordinates} candidates.
The majority ($\sim 80\%$) of $ZH$ candidates pass the trigger selection requiring events to contain at least one central  ($|\eta| \leq 1.0$) track of $p_T \ge 9\unit{GeV/}c$ matched to an electromagnetic energy (EM) cluster 
of $E_T \ge 18\unit{GeV}$ (a trigger electron) or at least one central track of $p_T \ge 20\unit{GeV/}c$ pointing to signals in the muon detectors (a trigger muon).
The remaining fraction of $ZH$ candidate events comes from newly included data  selected by a trigger that requires
two or more EM clusters of $E_T \ge 18\unit{GeV}$ and $| \eta | \le 3.6$ without requiring that the clusters are associated with tracks (trackless trigger).
Events are further required to contain a lepton pair that forms a $Z$ candidate with mass in the range $76 \le M_{ll} \le 106 \unit{GeV}/c^2$.
Pairs of central leptons forming $Z$ candidates must have opposite charge;  electrons in the forward ($|\eta| > 1.0$) acceptance of the detector may not have an associated track and no charge requirement is imposed.


We divide the $Z$ candidates into two categories based on signal-to-background ratio ($S/B$).
The search for the signal in these two categories is conducted separately to improve sensitivity to a $ZH$ signal.
The high-$S/B$ category includes $Z$ candidates formed from a trigger muon and a second muon candidate with $p_T \ge 10\unit{GeV/}c$, or a trigger electron paired with a second electron candidate formed from either a central EM cluster of $E_T \ge 10\unit{GeV}$ matched to a track of $p_T \ge 5\unit{GeV/}c$ or a forward EM cluster of $E_T \ge 18\unit{GeV}$. 
The low-$S/B$ category contains $Z$ candidates in events satisfying the trackless trigger only or formed from a trigger electron paired with an isolated central track with $p_T \ge 20\unit{GeV/}c$ pointing to an uninstrumented region of the calorimeter. 
The low-$S/B$ category is included for the first time in the search for $ZH$ production at CDF.

Higgs boson candidates are assembled from pairs of jets.
We form jets from energy deposits in the calorimeter using the {\sc jetclu} algorithm~\cite{Blazey:1999tm} with a cone radius of $0.4$.
We consider only jets well contained in the calorimeter, $|\eta| \leq 2.0$, and well separated from the $Z$-decay leptons, $\Delta R \geq 0.4$~\cite{Objects}.
Measured jet energies are corrected to account for $\eta$-dependent variations in detector response, calorimeter coverage, and the expected contribution from additional  $p\bar{p}$ pair interactions in the same event~\cite{Bhatti:2005ai}.
Events are required to have one jet with $E_T \ge 25\unit{GeV}$ and a second of $E_T \ge 15\unit{GeV}$.
We refer to the events containing a $Z$ boson candidate and two such jets as the PreTag sample; $b$~quark identification (described below) is applied to the PreTag sample to form our final analysis samples.
There are 11,806 (3,061) events in the high (low) $S/B$ PreTag data sample, wherein we expect $5.0\pm0.7$ 
($0.8\pm0.1$) $ZH$ signal events for $M_H = 115 \unit{GeV}/c^2$.   
Inclusion of the low-$S/B$ subsample increases the total signal yield in the PreTag sample by $16\%$.

The PreTag sample consists mainly of $Z$+light flavor (l.f.) jet (\textit{u,d,s,g}) events, with smaller contributions from 
$Z$+heavy flavor (h.f.) jet (\textit{c,b}), $t\overline{t}$, and diboson processes.
To reduce the $Z$+l.f.~ background, we look for $b$~jets in the event. 
We use two algorithms to identify (tag) $b$ jets:  one based on evidence for a decay displaced spatially from the $p\bar{p}$ interaction point (SV)~\cite{detector} and one based on track impact parameters with respect to the $p\bar{p}$ interaction point (JP)~\cite{Abulencia:2006kv}.
For the SV algorithm, there are two operating points: tight and loose~\cite{Neu:2006rs}.
The tight operating point has better l.f.-jet rejection (smaller mistag probability)  at the expense of reduced $b$-jet identification efficiency.

We select events in the PreTag sample using the $b$~tagging algorithms on the  jet pairs forming Higgs candidates.
We require the jet pairs to satisfy one of the following classifications, in order of precedence from highest to lowest ratio of $ZH$ signal to background:
a pair containing two SV-tight-tagged jets is classified as tight-double-tagged (TDT);
a pair consisting of one SV-loose-tagged jet and a second JP-tagged jet is classified as loose-double-tagged (LDT);
and a pair where only one jet has a SV-tight-tag is classified as single-tagged (ST).
When there are multiple  $H \rightarrow b\bar{b}$ candidates in an event, the jet pair with the highest $b$~tag class is selected (TDT $>$ LDT $>$ ST).  
This is the first time the JP algorithm has been used in the $ZH\rightarrow \ell^+ \ell^- b\bar{b}$ channel.
While the $b$~tag selection matches the signal selection efficiency (approximately 60\%) and background rejection rate (96\%) of previous efforts, 
the addition of the new LDT class increases sensitivity to a $ZH$ signal by $6\%$.
Due to differing background composition, the events are divided into independent subsamples based on $b$-tagging class.
With two $Z$ boson $S/B$ categories and three $b$-tagging classes, we form a total of six subsamples that we analyze for $ZH$ content.
We compare the $b$-tag data to a model of signal and backgrounds to estimate the signal content.
Signal, $t\overline{t}$, and diboson events are modeled with the {\sc pythia}~\cite{pythia} event generator.
Backgrounds from $Z$+h.f.~ processes are simulated at the quark level using {\sc alpgen}~\cite{alpgen}, then passed to {\sc pythia} for hadronization.
The detector response is modeled with a detailed detector simulation~\cite{detector}.
We estimate the background from $Z$+l.f.~ mistags using re-weighted PreTag data with weights reflecting the probability for a l.f.~ jet to be erroneously $b$~tagged.
A small fraction, much less than 1\%, of jets can be erroneously identified as electrons, resulting in a background of misidentified (misID) $Z\rightarrow ee$ candidates. 
The likelihood of jet-electron misidentification is measured in generic jet data and applied to all electron-jet and jet-jet pairings in the electron triggered samples to generate a model of misID $Z$ events.
The misID $Z\rightarrow\mu\mu$ background contribution is estimated from the events passing all selection requirements but containing like-charge muon pairs.
Predicted and observed event totals are listed in Table~\ref{table:TSel} for the $b$-tagged subsamples.

\begin{table*}
\renewcommand\arraystretch{1}
\begin{center}
\caption{
Comparison of the expected mean event totals for background and $ZH$ signal with the observed number of data events.
The totals are for full event selection, and systematic and statistical uncertainties are combined in quadrature.
Systematic uncertainties dominate.
The background composition is provided in the first five rows.
Each of the six $ZH$ subsamples is presented in a separate column.
}
\label{table:TSel}
\smallskip
\begin{tabular}{c  c c c   c@{\hspace{12pt}}    c c c}
 \hline\hline 
& \multicolumn{3}{c}{High $S/B$} &&\multicolumn{3}{c}{Low $S/B$} \\ \cline{2-4} \cline{6-8}
 Process  &{TDT} & {LDT} & {ST}  && {TDT} & {LDT} & {ST} \\
 \hline 
$t\bar{t}$	& $7.0\pm 1.5$        &$8 \pm 2$        &$ 17 \pm 4 $ &			&$ 2.9 \pm 0.6 $        &$ 3.2 \pm 0.8 $        &$ 8.9 \pm 1.9 $\\
Diboson	& 			$2.9\pm0.4$		&	$4\pm 1 $	&	$16\pm2$ 	&&		$0.5\pm 0.1 $		&		$ 0.6 \pm$0.1	&		$ 3.3 \pm 0.5 $		\\
$Z+\mathrm{h.f.} $	&	$18\pm7$	&	$30\pm 13 $&		$159\pm67$		&	&		$3.5\pm1.5$		&	$5.6\pm 2.4 $	&		$32 \pm14$	\\
$Z+\mathrm{l.f.}$	&	$ 0.9 \pm 0.3 $        &$ 9 \pm 3 $        &$ 152 \pm 23 $&&	$ 0.4 \pm 0.1 $        &$ 3.8 \pm 1.3 $        &$ 50\pm 7.6 $\\
misID $Z$	& 	 $0.7 \pm 0.3 $ &$ 2 \pm 1 $ &$ 22\pm 11$ & &	 $ 1.4 \pm 0.7 $   &$ 1.1 \pm 0.5 $  & $ 23 \pm 12 $\\
 \hline 
Total Bkg.  & 	$29 	\pm	8$&	$53 	\pm	14$ &	$366 \pm	72$ 	&	&	$9 \pm	2$	 &	$14 \pm	3$ &	$117 \pm20$ \\
$ZH (115 \unit{GeV}/c^{2})$  	     &$ 0.7 \pm 0.1 $        &$ 0.7 \pm 0.1 $        &$ 1.7 \pm 0.2 $&	 &$ 0.11\pm 0.01 $       &$ 0.11 \pm 0.03 $       &$ 0.28 \pm 0.05 $\\
\hline 				
Data	&  23 & 56 & 406 			&&$  12 $        &$  14 $        &$  116 $\\
\hline 
\hline
  \end{tabular}
\end{center}
\end{table*}

\begin{figure}[t]
\begin{center}
\includegraphics[width=1\columnwidth]{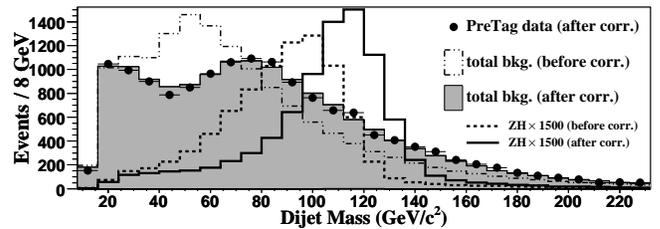} 
\caption{
The dijet invariant mass distribution of the two jets with the highest $E_T$ in the PreTag sample. 
 The distribution is shown for data after NN correction of jet energies.
 The dijet mass is shown for background before (dash-dotted line) and after (grey region) NN correction. 
 The dotted (solid) curves show the dijet mass for signal ($M_H = 115 \unit{GeV/}c^2$ and scaled by a factor of 1500) before (after) correction.
 }
\label{fig:Mjj}
\end{center}
\end{figure}

To improve the separation of $ZH$ from background, we utilize several multivariate techniques that use kinematic quantities as inputs.
The dijet mass $M_{jj}$ is one of the most useful quantities, with its separating power limited mainly by the jet-energy resolution.
In $ZH$ signal events with $Z\to\ell^+\ell^-$, incorrect measurement of jet energies results in apparent missing transverse energy $\missETvec$~\cite{MET}.
We correct jet energies, based upon the $\missETvec$, and thereby improve the resolution on the dijet invariant mass.
Jet-energy correction factors are computed by a NN trained to match measured jet energies to parton-level energies in $Z$+jets and signal events.  
In this analysis, this neural network is improved compared to the previous analysis~\cite{ZHllbb_2008} by utilizing additional input variables describing the recoil of the $Z$ boson. 
The corrected jet energies are used to recompute the Higgs candidate mass $M_{H}$, the $p_T$ of the jets, the $p_T$ of the Higgs candidate, the projection of $\missETvec$ onto the lower-$E_T$ Higgs jet, and the sphericity.
The effect of the NN corrections on the reconstructed $H\rightarrow b \bar{b}$ mass is shown in Fig.~\ref{fig:Mjj}.
In signal the resolution~\cite{RES} on $M_H$ is improved from about 18\% to 12\%.


\begin{figure*}[t]
\begin{center}
\includegraphics[width=2\columnwidth]{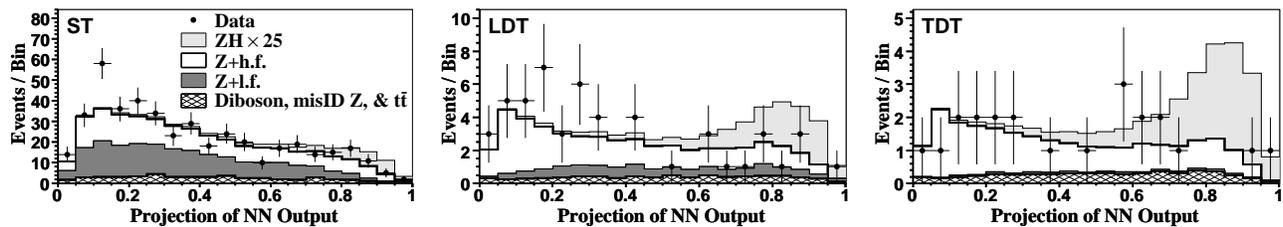}
\caption{
Projections of the two-dimensional neural network (NN$_{D}$) output onto the $x$-axis ($x$ and $y$ are defined in the text) for events in the $b$-tag categories ST, LDT, and TDT. 
Events with a NN$_{D}$ score of $y \ge 0.1$ are omitted to highlight the signal region.
The $ZH$ contribution is shown, multiplied by a factor of 25, for $M_H = 115 \unit{GeV/}c^2$.}
\label{fig:NN}
\end{center}
\end{figure*}

To exploit the combined signal-to-background discrimination power of event quantities and their correlations, we employ a neural network discriminant (NN$_{D}$) trained to simultaneously separate $ZH$, $t\bar{t}$, and $Z+\mathrm{jets}$ events. 
The NN$_{D}$ is configured to return values of $(x,y)=(1,0)$ for $ZH$ events, $(0,0)$ for $Z+\mathrm{jets}$, and $(1,1)$ for $t\bar{t}$.
A separate NN$_{D}$ is formed for each of the three $b$-tag classes.  
In addition to the quantities recomputed with corrected jet energies, the NN$_D$ inputs include:
$\missET$; MEPs for $ZH$, $t\bar{t}$, and $Z+\mathrm{jets}$ processes; the number of jets in the event; and the output of a $b$~jet identifying artificial neural network (NN$_b$)~\cite{KNN}.
The $ZH$ process MEP is computed~\cite{ZHLLBBME} by convolving the theoretical matrix element for $ZH$ production with detector resolution functions.
The resulting MEP reflects the likelihood that an event is signal.
We calculate two additional MEPs under the hypotheses that an event is $Z+\mathrm{jets}$ or $t\bar{t}$ to aid background rejection.  
The NN$_b$ augments the performance of the SV algorithm by isolating incorrectly 
$b$-tagged l.f.~ jets.
The addition of NN$_{b}$ as an input enhances the ability of the NN$_{D}$ to distinguish $ZH$ from $Z$+l.f., which constitutes 40\% of the total background in the ST class. 
One-dimensional projections of NN$_{D}$ output are shown in Fig.~\ref{fig:NN} for each $b$-tag class.

The effect of systematic uncertainties on the determination of signal content requires propagating uncertainties on NN$_{D}$ input quantities to the output distributions.
The dominant effect on the result comes from uncertainties on cross sections for background processes --- a $40\%$ uncertainty is assumed on the normalization of $Z+\mathrm{h.f.}$ samples~\cite{Campbell:2000bg}, $11.5\%$ for the diboson samples~\cite{CampbellEllis},
$20\%$ to account for the theoretical and experimental uncertainty on the $t\bar{t}$ cross section~\cite{PhysRevD.80.054009}, and $5\%$ for $ZH$ signal~\cite{Baglio:2010um}.
Uncertainties of $4\%$ (ST), $8\%$ (TDT), and $11\%$ (LDT) on the normalization of $b$-tagged samples are applied to account for different $b$-tag efficiencies in data and simulation.
Other uncertainties affecting sample normalizations include: $6\%$ on the integrated luminosity, $1\%$ on the trigger and lepton reconstruction efficiencies~\cite{lumi}, $1.5\%$ on the measurement of lepton energies, and a $50\%$ uncertainty on the total lepton misID estimate.
We include uncertainties on jet energies, the modeling of initial and final state radiation, and mistag event weighting as variations on the shape and normalization of the NN$_{D}$ output.

\begin{figure}[t]
\begin{center}
\includegraphics[width=1\columnwidth]{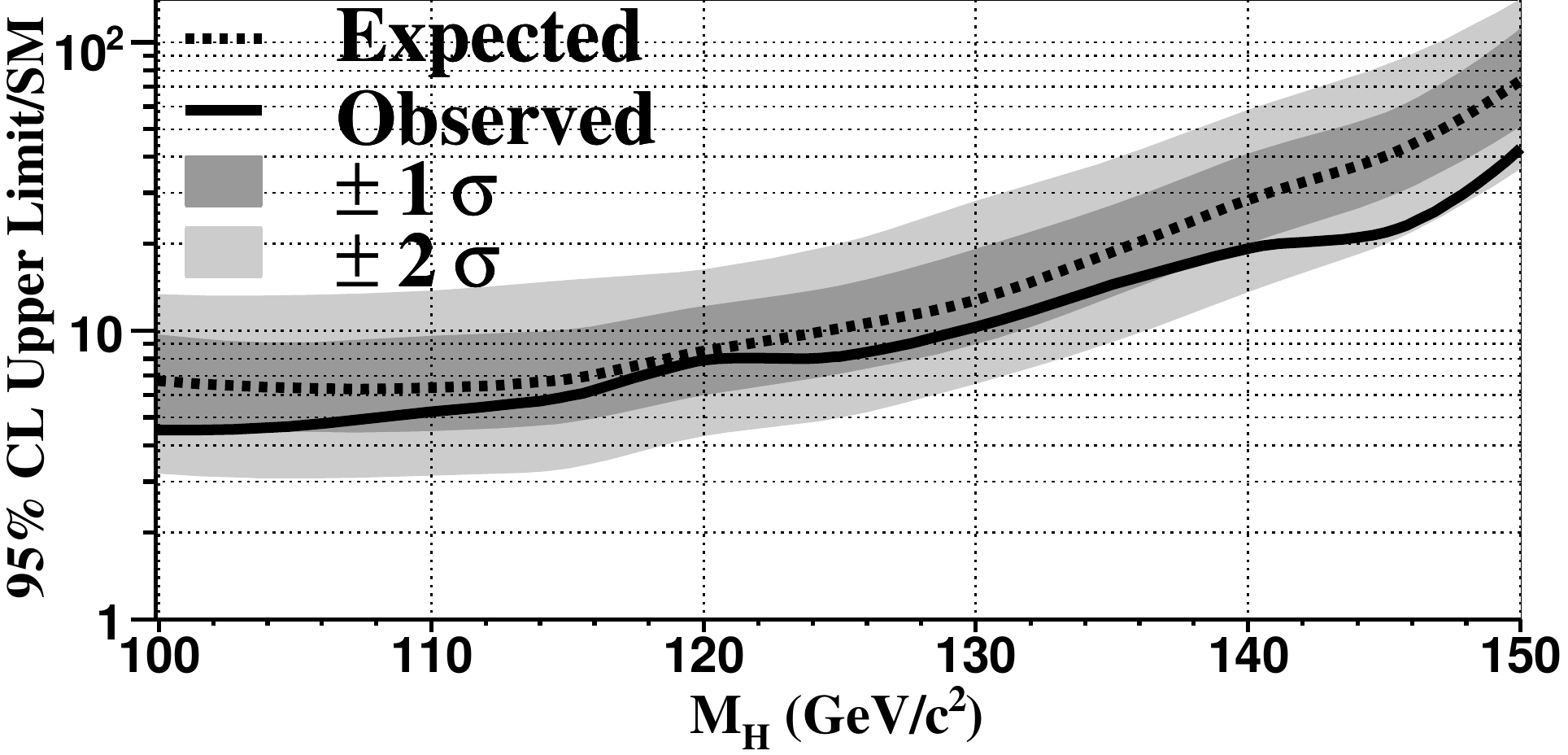}
\caption{The expected (dashed curve) and observed (solid curve) $ZH$ cross section upper limits divided by the SM cross section are shown as a function of the Higgs boson mass.
The dark (light) band represents the $\pm1\sigma$ ($\pm2\sigma$) expected limit range. }
\label{fig:limit}
\end{center}
\end{figure}

We calculate a limit on the associated production of a SM Higgs boson and $Z$ boson based on the comparison of the full NN$_D$ output of the $b$-tagged data to expectations for signal and background.
We consider eleven Higgs mass hypotheses between $100$ and $150 \unit{GeV}/c^2$.
We use a Bayesian algorithm~\cite{PDG2010} with a flat prior in the production cross section, integrating over the priors for the systematic uncertainties, incorporating correlated rate and shape uncertainties, and uncorrelated bin-by-bin statistical uncertainties \cite{aaltonen:012002}.
Systematic uncertainties reduce the sensitivity of this search by 16\%.
The expected 95\% credibility level (C.L.) limits are calculated assuming no signal, based on expected backgrounds only, as the median of 1000 simulated experiments.
The $\pm1\sigma$ and $\pm2\sigma$ expected limits are derived from the distribution of the simulation limits at the $16^{\mathrm{th}}$, $84^{\mathrm{th}}$, $2^{\mathrm{nd}}$, and $98^{\mathrm{th}}$ percentiles of the distribution, respectively.
The observed 95\% C.L. on the $ZH$ cross section are calculated from the $b$-tagged data, and displayed in Fig.~\ref{fig:limit} and summarized in Table~\ref{table:lim1}.

\begin{table}[htdp]	
\caption{Expected (Exp.) and observed (Obs.) $95\%$ C.L. upper limits on the $ZH$ production cross section times the branching ratio for $H\rightarrow b \bar{b}$ normalized to the SM value for Higgs masses between $100$ and $150\unit{GeV}/c^2$.
The assumed $ZH$ cross section and branching fraction for $H\rightarrow b \bar{b}$ are $0.11\unit{pb}$~\cite{fixBrein:2003wg,fixCiccolini:2003jy} and 0.73~\cite{BBMAIN} for a $115\unit{GeV/}c^2$ Higgs boson.}
 \renewcommand\arraystretch{1}
\begin{center}											
\begin{tabular}{c c c c c c c c c c c c}										
 \hline											
 \hline											
$M_H$ & 100 & 105 & 110 & 115 & 120 & 125 & 130 & 135 & 140 & 145 & 150 \\
\hline
Exp. & 6.7 & 6.4 & 6.3 & 6.8 & 8.5 & 10. & 13 & 19 & 29 &  45 & 74 \\
Obs. & 4.5 & 4.6 & 5.3 & 5.9 & 7.9 & 8.1 & 10. & 14 & 19 &  24 & 43 \\

\hline		
 \hline											
  \end{tabular}
 \end{center}											
\label{table:lim1}
\end{table}
%
In conclusion, we have searched for the SM Higgs boson produced in association with a $Z$ boson, where $Z \rightarrow \ell^+ \ell^-$ and $H \rightarrow b \bar{b}$, finding no significant evidence for the process.
We set 95\% C.L. upper limits on the $ZH$ production cross section times the $H\rightarrow b\bar{b}$ branching ratio for Higgs boson masses between
$100$ and  $150\unit{GeV/}c^2$.
For a Higgs boson mass of $115\unit{GeV}/c^2$ we set (expect) a $95\%$ C.L. upper limit of $5.9$ ($6.8$) times the standard model prediction.
This result is an important step forward in the search for the Higgs boson and the source of EWSB, improving upon the previous observed (expected) limits in this channel by factors of 2.2 to 3.7 (1.9 to 2.4).

%
We thank the Fermilab staff and the technical staffs of the participating institutions for their vital contributions. This work was supported by the U.S. Department of Energy and National Science Foundation; the Italian Istituto Nazionale di Fisica Nucleare; the Ministry of Education, Culture, Sports, Science and Technology of Japan; the Natural Sciences and Engineering Research Council of Canada; the National Science Council of the Republic of China; the Swiss National Science Foundation; the A.P. Sloan Foundation; the Bundesministerium f\"ur Bildung und Forschung, Germany; the World Class University Program, the National Research Foundation of Korea; the Science and Technology Facilities Council and the Royal Society, UK; the Institut National de Physique Nucleaire et Physique des Particules/CNRS; the Russian Foundation for Basic Research; the Ministerio de Ciencia e Innovaci\'{o}n, and Programa Consolider-Ingenio 2010, Spain; the Slovak R\&D Agency; and the Academy of Finland.

\bibliography{ZHllbb}

\begin{thebibliography}{31}
\bibitem[{coo()}]{coordinates}
\bibinfo{note}{We use a cylindrical coordinate system with $z$ along the proton
  beam direction, $r$ the perpendicular radius from the central axis of the
  detector, and $\phi$ the azimuthal angle. For $\theta$ the polar angle from
  the proton beam, we define $\eta = -\ln\tan(\theta/2)$, 
  transverse momentum $p_T = p \sin\theta$ and transverse energy $E_T = E
  \sin\theta$.}
\bibitem[{Obj()}]{Objects}
\bibinfo{note}{The $\Delta R$ between two objects is given by
  $\sqrt{\Delta\eta^2 + \Delta\phi^2}$}.
\bibitem[{MET()}]{MET}
\bibinfo{note}{The missing $E_T$ ($\not\!\!\vec{E}_T$) is defined by the sum
  over calorimeter towers: $\not\!\!\vec{E}_T = - \sum_{i} E_T^i \hat{n}_i$,
  where $i =$ calorimeter tower number with $|\eta| < 3.6$, $\hat{n}_i$ is a
  unit vector perpendicular to the beam axis and pointing at the i$^{th}$
  calorimeter tower. We also define $\not\!\!E_T=|{\not\!\!\vec{E}_T}|$.}
\bibitem[{RES()}]{RES}
\bibinfo{note}{We define dijet mass resolution as the standard deviation of the
  dijet mass distribution divided by its mean.}
\end{thebibliography}

\end{document}